\newif\ifplain
\plaintrue

\ifplain
\documentclass[10pt,a4paper]{article}
\fi

\usepackage[usenames,dvipsnames]{xcolor} 

\usepackage[linesnumbered,ruled]{algorithm2e}
	\definecolor{comment}{rgb}{0,0.25,0}
	\SetAlgorithmName{Algorithm}{Algorithm}{List of Algorithms}
	\SetAlgoProcName{Procedure}{Procedure}
	\SetAlgoFuncName{Function}{Function}
	\SetAlFnt{\small}
	\SetAlCapFnt{\small}
	\SetAlCapNameFnt{\small}
	\SetAlCapHSkip{0pt}
	\SetProcFnt{\small}
	\SetProcNameFnt{\small}
	\SetProcArgFnt{\small}

	\SetProcNameSty{textsc}
	\SetKw{KwAnd}{and}
	\SetKw{KwBreak}{break}
	\SetKw{KwDownTo}{downto}
	\SetKw{KwExit}{exit}
	\SetKw{KwOr}{or}
	\SetKw{KwReturn}{return}
	\SetKwFor{ForAll}{for all}{do}{end forall}
	\SetKwFor{ForAllPar}{for all}{do in parallel}{end forallpar}

\usepackage{amsmath}
\usepackage{amssymb}
\usepackage{booktabs}
\usepackage{cite}

\usepackage{graphicx}

\usepackage[hidelinks]{hyperref}

\usepackage{mathtools}
	\mathtoolsset{showonlyrefs=true} 

\usepackage{pgfplots}
\pgfplotsset{compat=newest}
\pgfplotsset{
	tick label style = {font=\sansmath\sffamily\small},
	every axis label = {font=\sansmath\sffamily\small},
	legend style = {font=\sffamily\small},
	label style = {font=\sansmath\sffamily\small},
	title style = {font=\sansmath\sffamily\small},
}
\pgfplotscreateplotcyclelist{cyclelist1}{%
	darkgray,mark=*,solid\\%
	blue!50!cyan,mark=square*,solid\\%
	purple,every mark/.append style={scale=1.3},mark=triangle*,solid\\%
	orange,every mark/.append style={scale=1.3},mark=diamond*,solid\\%
	black!50!blue,mark=*,dashed\\%
	red,mark=square*,dashed\\%
	black!50!green,every mark/.append style={scale=1.3},mark=triangle*,dashed\\%
	magenta,every mark/.append style={scale=1.3},mark=diamond*,dashed\\%
}
\pgfplotscreateplotcyclelist{cyclelist11}{%
	darkgray,thick,mark=*,solid\\%
	blue!50!cyan,thick,mark=square*,dashed\\%
	purple,thick,every mark/.append style={scale=1.3},mark=triangle*,dashdotted\\%
	black!20!orange,thick,every mark/.append style={scale=1.3},mark=diamond*,dashdotdotted\\%
}
\pgfplotscreateplotcyclelist{cyclelist2}{%
	black,mark=*,solid\\%
	blue,mark=square*,solid\\%
	red,every mark/.append style={scale=1.3},mark=triangle*,solid\\%
	orange,every mark/.append style={scale=1.3},mark=diamond*,solid\\%
}
\usepgfplotslibrary{external}
\tikzsetexternalprefix{images/tikz/}
\tikzsetfigurename{fig}
\expandafter\gdef\csname c@tikzext@no@\pgfkeysvalueof{/tikz/external/figure name}\endcsname{1}%

\usepackage{pgfplotstable}
\usepackage{sansmath}

\usepackage[final,notref,notcite,color]{showkeys}
\definecolor{labelkey}{rgb}{1,0,1}

\usepackage{subfig}
\usepackage{xspace}

\ifplain
\usepackage[margin=25mm,includefoot]{geometry}
\usepackage[shortlabels]{enumitem}
\setlist{itemsep=0.3ex,parsep=0ex}
\fi

\providecommand{\nnz}{\ensuremath{\mathit{nnz}}\xspace}
\providecommand{\var}[1]{\ensuremath{\mathit{#1}}\xspace}

\begin{document}

\title{Loading Large Sparse Matrices Stored in Files in the Adaptive-Blocking Hierarchical Storage Format}

\author{%
Daniel Langr\footnote{E-mail: \href{mailto:langrd@fit.cvut.cz}{langrd@fit.cvut.cz}} \ and Ivan \v{S}ime\v{c}ek \ and Pavel Tvrd{\' \i}k \\
\small Czech Technical University in Prague \\
\small Department of Computer Systems \\
\small Faculty of Information Technology \\
\small Th\'{a}kurova 9, 160 00, Praha, Czech Republic \\
}


\maketitle

\begin{abstract}

The parallel algorithm for loading large sparse matrices from files into distributed memories of high performance computing (HPC) systems is presented. This algorithm was designed specially for matrices stored in files in the space-efficient adaptive-blocking hierarchical storage format (ABHSF). The algorithm can be used even if matrix storing and loading procedures use a different number of processes, different matrix-processes mapping, or different in-memory storage format. The file format based on the utilization of the HDF5 library is described as well. Finally, the presented experimental study evaluates the proposed algorithm empirically.


\end{abstract}

\section{Introduction}
\label{sec:intro}


Large sparse matrices emerge frequently in high performance computing (HPC) %
applications. Sometimes, these matrices need to be stored to a file system and then loaded back, e.g., when the checkpointing-restart (C/R) resiliency technique is applied. The runtime of the store/load process is generally proportional to the amount of data processed by the I/O subsystem. Common in-memory sparse matrix storage formats used for computations are typically not much space efficient. We have shown that when matrices need to be saved to a file system, it pays off to convert them into some highly space-efficient format, such as the ABHSF~\cite{Langr:2013a}. However, this approach brings some runtime overhead into the I/O processes since matrices need to be converted between the in-memory format and the ABHSF on the fly. 

We have previously introduced and experimentally evaluated algorithms for conversion of sparse matrices to the ABHSF from the most commonly used formats---the \emph{coordinate} (COO) and \emph{compressed sparse rows} (CSR) formats~\cite{Saad:2003,Barrett:1994}. In this article, we focus on the successive step of loading matrices back from a file system to memory. 

Modern HPC systems are based on hybrid distributed/shared memory architectures. They consists of shared-memory nodes connected by fast networks. Matrices emerging in these systems thus need to be mapped to separate address spaces of different application (typically MPI) processes. Each process then takes care of some portion of matrix nonzero elements, which are in its address space present in memory in a sparse storage format. The conditions under which are matrices treated are given by 
\begin{enumerate}
\item the number of application processes,
\item the particular mapping of matrix nonzero elements to these processes,
\item the sparse storage format used for storing the to-process mapped elements in its address space.
\end{enumerate}
We further call these conditions simply the \emph{configuration}.

When matrices need to be loaded from a file system to memory, the configuration can be either same or different compared to the configuration used during the storage process. We consider both these options within this text and study their influence to the performance of the matrix loading processes.

\section{Data and File Structures}

Let $A=(a_{i,j})$ be an $m\times n$ matrix with \nnz nonzero elements. Let further $A$ be stored in HDF5-based files in the ABHSF sparse storage format. The process of creating these files (matrix storage) including the storage algorithms was described by Langr et al.~\cite{Langr:2013a}. For all file operations, we use the HDF5 library~\cite{HDFGroup:2013} for both serial and parallel I/O; it is a de-facto standard for intensive I/O operations in HPC. Due to microbenchmarking performed on various modern HPC system, we chose a single-file-per-process strategy for storing matrices (in contrast to shared-file strategy where a single file is shared by all processes); it generally provided higher I/O performance in our measurements. This strategy means that during matrix storage, each process stores its local nonzero elements into a separate file independent of files accessed by other processes. All these files are stored in the directory called \texttt{matrix} in a HDF5-based file called \texttt{matrix-}$k$\texttt{.h5spm}, where $k$ denotes a process number (MPI process rank).

Assume that prior to the matrix storage, $A$ was treated by $P$ processes denoted by $p_1,\ldots,p_P$. Let $\mathcal{A}^{(k)}$ denote the set of nonzero elements of $A$ stored in the address space of process $p_k$. The nonzero elements of $A$ are distributed among processes such that all the nonzero elements treated by process $p_k$ falls to a submatrix of $A$ that starts at row $r^{(k)}$ and column $c^{(k)}$ and has the size $m^{(k)}\times n^{(k)}$. Thus,
\begin{align}
r^{(k)} &= \min_{a_{i,j}\in \mathcal{A}^{(k)}} i, &
m^{(k)} &= \max_{a_{i,j}\in \mathcal{A}^{(k)}} i - r^{(k)} + 1, \\
c^{(k)} &= \min_{a_{i,j}\in \mathcal{A}^{(k)}} j &
n^{(k)} &= \max_{a_{i,j}\in \mathcal{A}^{(k)}} j - c^{(k)} + 1.
\end{align}
%
Then, for all the nonzero elements $a_{i,j}\in \mathcal{A}^{(k)}$ treated by process $p_k$ holds $r^{(k)} \leq i < r^{(k)}+m^{(k)}$ and $c^{(k)} \leq j < c^{(k)}+n^{(k)}$. In most general case, $r^{(k)}=c^{(k)}=1$, $m^{(k)}=m$ and $n^{(k)}=n$ might hold for each process $p_k$. However, in practice, one- or two-dimensional partitioning schemes are most commonly used for matrix-processes mapping problem due to optimization of communication needed during sparse matrix-vector multiplication (SpMV) operation; see, e.g., \cite{Catalyurek:2010} for a survey of these schemes.

In contrary of the common mathematical notation used above, we further consider 0-based indexing for data structures and algorithm pseudocodes. The ABHSF is based on partitioning of the local (per-process) submatrix to a fixed blocks of sizes $s\times s$; this format is described in detail by Langr et al.~\cite{Langr:2012b}. The structure of the \texttt{matrix-}$k$\texttt{.h5spm} file is as follows:
\begin{tabbing}
\hspace*{1em} \= \hspace*{8em} \= \kill
\textbf{structure} \var{abhsf} $\mathrel{\mathop:}= \{$ \\
\> \var{m}:           \> number of rows $m$; \\
\> \var{n}:           \> number of columns $n$; \\
\> \var{z}:           \> number of nonzero elements \nnz; \\
\> \var{m\_local}:    \> number of local rows $m^{(k)}$; \\
\> \var{n\_local}:    \> number of local columns $n^{(k)}$; \\
\> \var{z\_local}:    \> number of local nonzero elements $\var{nnz}^{(k)}=|\mathcal{A}^{(k)}|$; \\
\> \var{m\_offset}:   \> first row of local submatrix $r^{(k)}$; \\
\> \var{n\_offset}:   \> first column of local submatrix $c^{(k)}$; \\
\> \var{block\_size}: \> block size $s$; \\
\> \var{blocks}:      \> number of nonzero blocks of local submatrix; \\
\> \var{schemes[]}:   \> scheme tags for nonzero blocks (COO, CSR, bitmap, dense); \\
\> \var{zetas[]}:     \> number of nonzero elements of nonzero blocks; \\
\> \var{brows[]}:     \> block row indexes of nonzero blocks; \\
\> \var{bcols[]}:     \> block column indexes of nonzero blocks; \\
\> \var{coo\_lrows[]}:     \> in-block row indexes of nonzero elements for COO blocks; \\
\> \var{coo\_lcols[]}:     \> in-block column indexes of nonzero elements for COO blocks; \\
\> \var{coo\_vals[]}:      \> in-block values of nonzero elements for COO blocks; \\
\> \var{csr\_lcolinds[]}:  \> in-block column indexes of nonzero elements for CSR blocks; \\
\> \var{csr\_rowptrs[]}:   \> in-block offsets of rows data for CSR blocks; \\
\> \var{csr\_vals[]}:      \> in-block values of nonzero elements for CSR blocks; \\
\> \var{bitmap\_bitmap[]}: \> bitmap structure of nonzero elements for bitmap blocks; \\
\> \var{bitmap\_vals[]}:   \> in-block values of nonzero elements for bitmap blocks; \\
\> \var{dense\_vals[]}:    \> in-block values of all elements for dense blocks; \\
$\}$.
\end{tabbing}

The data name accompanied with $[]$ denote HDF5 datasets (generally arrays). Other data names denote HDF5 attributes (generally simple variables). 

Let \var{csr} be a data structure that represents a matrix in the CSR storage format in a computer memory of process $p^{(k)}$ defined as follows:
\begin{tabbing}
\hspace*{1em} \= \hspace*{6em} \= \kill
\textbf{structure} \var{csr} $\mathrel{\mathop:}= \{$ \\
\> \var{m}:           \> number of rows $m$; \\
\> \var{n}:           \> number of columns $n$; \\
\> \var{z}:           \> number of nonzero elements \nnz; \\
\> \var{m\_local}:    \> number of local rows $m^{(k)}$; \\
\> \var{n\_local}:    \> number of local columns $n^{(k)}$; \\
\> \var{z\_local}:    \> number of local nonzero elements $\var{nnz}^{(k)}=\mathcal{A}^{(k)}$; \\
\> \var{m\_offset}:   \> first row of local submatrix $r^{(k)}$; \\
\> \var{n\_offset}:   \> first column of local submatrix $c^{(k)}$; \\
\> \var{vals[]}:      \> values of nonzero elements; \\
\> \var{colinds[]}:   \> column indexes of nonzero elements; \\
\> \var{rowptrs[]}:   \> indexes of data for individual rows; \\
$\}$.
\end{tabbing}

Let \var{element\_t} be a auxiliary data structure representing a single matrix nonzero element defined as follows:
\begin{tabbing}
\hspace*{1em} \= \hspace*{4em} \= \kill
\textbf{structure} \var{element\_t} $\mathrel{\mathop:}= \{$ \\
\> \var{row}:	\> row index; \\
\> \var{col}:	\> column index; \\
\> \var{val}:   \> value; \\
$\}$.
\end{tabbing}

\section{Algorithms}

The pseudocode for loading matrices from files stored in the ABHSF into computer memory is presented by \autoref{alg:direct}--\ref{alg:loadblockdense}. In memory, the loaded local matrix nonzero elements are as output stored in the CSR format. The algorithms can be easily adapted for the COO format as well; COO is simpler and more generic than CSR (another option is to convert elements from CSR to COO afterwards, such conversion is straightforward). 

Due to the length of the pseudocode, the algorithm is recursively split into several procedures. However, we assume that all variables and arrays have a global scope, i.e., they are directly accessible inside procedures as well (without passing them as parameters).

The presented pseudocode works for the same configuration that was used for matrix storage. However, it can be adapted even for situations where different store/load configuration is needed. Let $\mathcal{M}(i,j)$ be the id/rank of a process that should have, in its local memory, a matrix elements $a_{i,j}$ after the loading procedure. The algorithm for different storing and loading configurations differ from \autoref{alg:direct}--\ref{alg:loadblockdense} only slightly, thus we do not present it as a complete pseudocode. The changes consists of the following steps:
\begin{enumerate}
\item The presented algorithm need to be encapsulated with the outer loop, in which \emph{all} processes read \emph{all} stored files.
\item The read nonzero elements are stored into memory of process $k$ only if $\mathcal{M}(i,j)=k$.
\end{enumerate}

This adaption covers an arbitrary mapping of matrix nonzero elements to processes (given by the \emph{mapping function} $\mathcal{M}$). Moreover, it also covers situations, where a different numbers of processes are used during storing and loading procedures. When a different in-memory storage format is finally required, the most straightforward way is to store elements in COO, sort them accordingly, and finally convert into the desired format. Since there many such formats exists in practice, such conversion is beyond the scope of this text.

\begin{algorithm}[p]
\DontPrintSemicolon
\KwIn{\var{abhsf}}
\KwOut{\var{csr}}
\KwData{$s$, $Z$, \var{elements}, \var{zeta}, \var{brow}, \var{bcol}, \var{last\_brow}, $k$, \var{row}, $l$}
\BlankLine
$\var{csr.m}\gets \var{abhsf.m}$ \;
$\var{csr.n}\gets \var{abhsf.n}$ \;
$\var{csr.z}\gets \var{abhsf.z}$ \;
$\var{csr.m\_local}\gets \var{abhsf.m\_local}$ \;
$\var{csr.n\_local}\gets \var{abhsf.n\_local}$ \;
$\var{csr.z\_local}\gets \var{abhsf.z\_local}$ \;
$\var{csr.m\_offset}\gets \var{abhsf.m\_offset}$ \;
$\var{csr.n\_offset}\gets \var{abhsf.n\_offset}$ \;
$s\gets \var{abhsf.block\_size}$ \;
$Z \gets \var{abhsf.blocks}$ \;
$\var{elements} \gets $ empty dynamic array of \var{element\_t} types \; 
\tcp{initial read of datasets entries:}
$\var{scheme} \gets \var{abhsf.schemes[0]}$ \;
$\var{zeta} \gets \var{abhsf.zetas[0]}$ \;
$\var{brow} \gets \var{abhsf.brows[0]}$ \;
$\var{bcol} \gets \var{abhsf.bcols[0]}$ \;
$\var{last\_brow} \gets 0$ \;
\For{$k \gets 0$ \KwTo $Z-1$}{
  call \textsc{LoadBlock} \;
  \If{$k < Z - 1$}{
    \tcp{next block:}
    $\var{scheme} \gets \var{abhsf.schemes[k+1]}$ \;
    $\var{zeta} \gets \var{abhsf.zetas[k+1]}$ \;
    $\var{brow} \gets \var{abhsf.brows[k+1]}$ \;
    $\var{bcol} \gets \var{abhsf.bcols[k+1]}$ \;
    \tcp{process block row if needed:}
    \If{$\var{brow} \neq \var{last\_brow}$ \KwAnd $k = Z-1$}{
      \lIf{\var{elements} $\mathrm{contains\ at\ least\ 2\ entries}$}{sort \var{elements} lexicographically} 
      \If{\var{elements} $\mathrm{contains\ at\ least\ 1\ entries}$}{
        $\var{row} \gets \var{brow} \times s$ \;
        \For{$l \gets 0$ \KwTo $\mathrm{size\ of} \var{elements} -1$}{
          \While{$\var{row} < \var{elements[l].row}$}{
            append $l$ to $\var{csr.rowptrs[]}$ \;
            $\var{row} \gets \var{row} + 1$ \;
          }
          append $\var{elements[l].col}$ to $\var{csr.colinds[]}$ \;
          append $\var{elements[l].val}$ to $\var{csr.vals[]}$ \;
          \While{$\var{row} < (\var{brow} + 1) \times s$}{
            append size of \var{elements} to $\var{csr.rowptrs[]}$ \;
            $\var{row} \gets \var{row} + 1$ \;
          }
          empty \var{elements} array \;
        }
        $\var{last\_brow} \gets \var{brow}$ \;
      }
    }
  }
}
\caption{Loading of matrices from files in the ABHSF into memory.}
\label{alg:direct}
\end{algorithm}

\begin{algorithm}[t]
\DontPrintSemicolon
\uIf{$\var{scheme} = \mathrm{COO}$}{
  call \textsc{LoadBlockCOO} \;
}
\uElseIf{$\var{scheme} = \mathrm{CSR}$}{
  call \textsc{LoadBlockCSR} \;
}
\uElseIf{$\var{scheme} = \mathrm{bitmap}$}{
  call \textsc{LoadBlockBitmap} \;
}
\uElseIf{$\var{scheme} = \mathrm{dense}$}{
  call \textsc{LoadBlockDense} \;
}
\Else{
  raise error (wrong scheme tag) \;
}
\caption{Procedure \textsc{LoadBlock}}
\label{alg:loadblock}
\end{algorithm}

\begin{algorithm}[t]
\DontPrintSemicolon
\KwData{$l$, \var{lrow}, \var{lcol}, \var{element}}
\BlankLine
\For{$l \gets 0$ \KwTo $\var{zeta} - 1$}{
  $\var{lrow} \gets$ next value from \var{abhsf.coo\_lrows[]} \;
  $\var{lcol} \gets$ next value from \var{abhsf.coo\_lcols[]} \;
  $\var{element} \gets$ variable of \var{element\_t} type \;
  $\var{element.row} \gets \var{lrow} + \var{brow} \times s$ \;
  $\var{element.col} \gets \var{lcol} + \var{bcol} \times s$ \;
  $\var{element.val} \gets$ next value from \var{abhsf.coo\_vals[]} \;
  append \var{element} into \var{elements} array \;
}
\caption{Procedure \textsc{LoadBlockCOO}}
\label{alg:loadblockcoo}
\end{algorithm}

\begin{algorithm}[t]
\DontPrintSemicolon
\KwData{\var{rowptrs\_1}, \var{lrow}, \var{rowptrs\_2}, \var{rowptr}, \var{lcol}, \var{element}}
\BlankLine
$\var{rowptrs\_1} \gets$ next value from \var{abhsf.csr\_rowptrs[]} \;
\For{$\var{lrow} \gets 0$ \KwTo $s - 1$}{
  $\var{rowptrs\_2} \gets$ next value from \var{abhsf.csr\_rowptrs[]} \;
  \For{$\var{rowptr} \gets \var{rowptrs\_1}$ \KwTo $\var{rowptrs\_2 - 1}$}{
    $\var{lcol} \gets$ next value from \var{abhsf.csr\_lcolinds[]} \;
    $\var{element} \gets$ variable of \var{element\_t} type \;
    $\var{element.row} \gets \var{lrow} + \var{brow} \times s$ \;
    $\var{element.col} \gets \var{lcol} + \var{bcol} \times s$ \;
    $\var{element.val} \gets$ next value from \var{abhsf.csr\_vals[]} \;
    append \var{element} into \var{elements} array \;
  }
  $\var{rowptrs\_1} \gets \var{rowptrs\_2}$ \;
}
\caption{Procedure \textsc{LoadBlockCSR}}
\label{alg:loadblockcsr}
\end{algorithm}

\begin{algorithm}[t]
\DontPrintSemicolon
\KwData{\var{bit}, \var{lrow}, \var{lcol}, \var{byte}, \var{element}}
\BlankLine
$\var{bit} \gets 8$ \;
\For{$\var{lrow} \gets 0$ \KwTo $s - 1$}{
  \For{$\var{lcol} \gets 0$ \KwTo $s - 1$}{
    \If{$\var{bit} > 7$}{
      $\var{byte} \gets$ next value from \var{abhsf.bitmap\_bitmap[]} \;
      $\var{bit} \gets 0$ \;
    }
    \If{$\mathrm{least\ significant\ bit\ in\ } \var{byte} = 1$}{
      $\var{element} \gets$ variable of \var{element\_t} type \;
      $\var{element.row} \gets \var{lrow} + \var{brow} \times s$ \;
      $\var{element.col} \gets \var{lcol} + \var{bcol} \times s$ \;
      $\var{element.val} \gets$ next value from \var{abhsf.bitmap\_vals[]} \;
      append \var{element} into \var{elements} array \;
    }
    $\var{byte} \gets \var{byte}$ shifted right of 1 bit \;
    $\var{bit} \gets \var{bit} + 1$ \;
  }
}
\caption{Procedure \textsc{LoadBlockBitmap}}
\label{alg:loadblockbitmap}
\end{algorithm}

\begin{algorithm}[t]
\DontPrintSemicolon
\KwData{\var{lrow}, \var{lcol}, \var{val}, \var{element}}
\BlankLine
\For{$\var{lrow} \gets 0$ \KwTo $s - 1$}{
  \For{$\var{lcol} \gets 0$ \KwTo $s - 1$}{
    $\var{val} \gets$ next value from \var{abhsf.dense\_vals[]} \;
    \If{$\var{val} \neq 0$}{
      $\var{element} \gets$ variable of \var{element\_t} type \;
      $\var{element.row} \gets \var{lrow} + \var{brow} \times s$ \;
      $\var{element.col} \gets \var{lcol} + \var{bcol} \times s$ \;
      $\var{element.val} \gets \var{val}$ \;
      append \var{element} into \var{elements} array \;
    }
  }
}
\caption{Procedure \textsc{LoadBlockDense}}
\label{alg:loadblockdense}
\end{algorithm}

\section{Experiments}

We have performed experiments with the experimental MPI/C++ implementation of the proposed algorithm using the Anselm HPC system operated by IT4Innovations, located in Ostrava, Czech Republic. This system is a Bullx Linux cluster consisting of 3.3k Intel Sandy Bridge cores, 15 TB of memory, an Infiniband interconnect and a Lustre parallel file system.

As a source of sparse matrices, we used the scalable parallel generator of matrices based on enlargement of small ``seed'' matrices by a Kronecker product operation, see~\cite{Langr:2013b} for details. As a seed matrix, we used the real unsymmetric square matrix \textsf{cage12} with 130k or rows and columns and 2M of nonzero elements. This matrix was enlarged so that the local per-process matrix part occupied 256 GB of memory using the COO format, double precision representation of element values, and 32 bit row and column indexes. 

As for configuration, the enlarged matrix was mapped to processes in a row-wise manner, i.e, each process took care or a contiguous chunk of rows such that the amortized number of nonzero elements treated by each process was the same. For matrix storage, we used 60 MPI processes.

We measured the loading times of the following cases:
\begin{enumerate}
\item the same loading configuration was used as within storage procedure (i.e., 60 MPI processes and balanced row-wise mapping);
\item a different number of processes and a regular column-wise mapping (same amortized number of columns per process) was used.
\end{enumerate}
The second test case (different configurations) were measured for two different HDF5 parallel I/O strategies: \emph{indepedent} and \emph{collective}; see~\cite{HDFGroup:2013} for details. Recall that in this case, all processes read all files, thus all processes access each file at once.

\begin{figure}[t]
\centering
\includegraphics{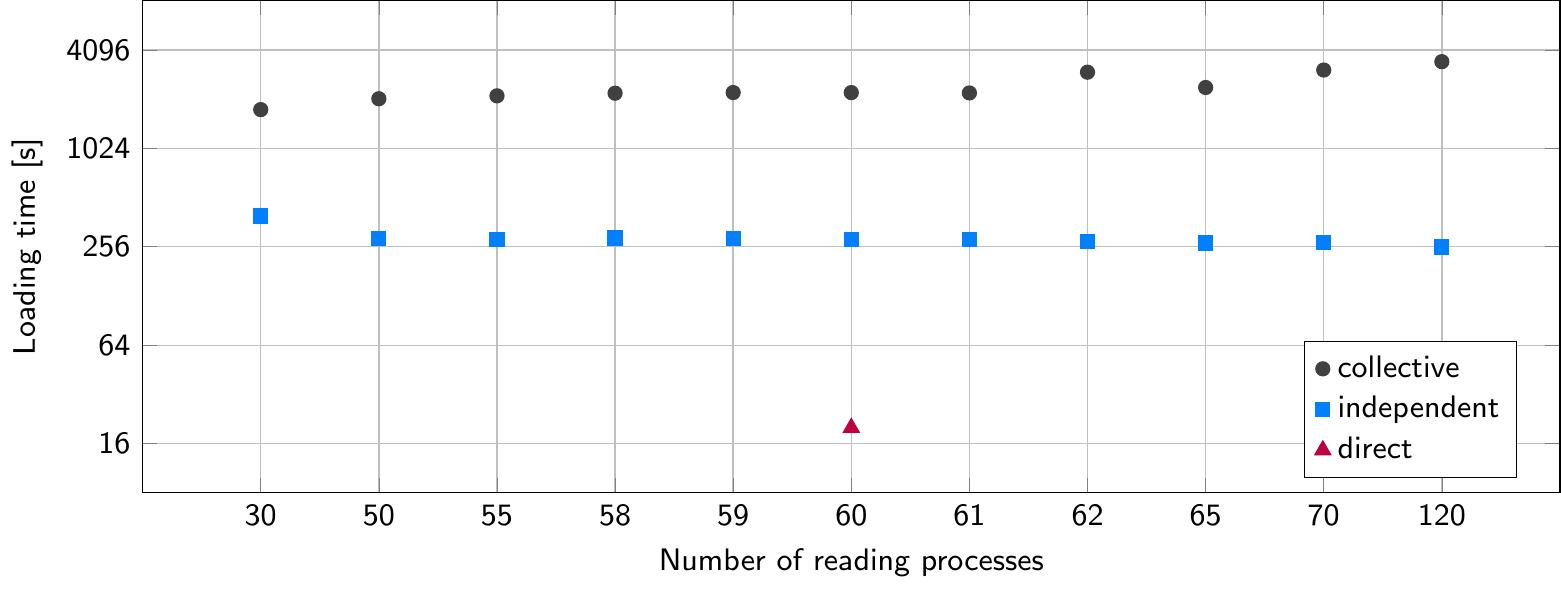}
\caption{Measured time of the process of loading matrices from the file system to memory for different configurations.}
\label{fig:results}
\end{figure}

The measured results are shown in \autoref{fig:results}. We can clearly see that when the storing and loading configurations match, the loading time is the lowest. Such a result was expected, since the overall amount of data processed by the I/O subsystem is the lowest as well. As for different configurations, the independent HDF5 strategy resulted in considerably lower loading times than the collective strategy. Moreover, these loading times were almost independent of the number of reading processes. We can also observe that these times are much lower than the loading time for the same configurations multiplied by the number of processes, which is proportional to the amount of processed data.

\pgfplotstableread{
processes collective independent direct 
30 1766.39 394.961 0
50 2058.73 286.18 0
55 2144.03 284.86 0
58 2225.29 289.376 0
59 2245.54 286.73 0
60 2244.9 284.733 20.0173                             
61 2233.36 285.016 0
62 2991.84 275.145 0
65 2411.73 270.195 0
70 3084.63 271.736 0
120 3470.52 255.111 0
}{\results}

\section{Conclusions}

We have presented an algorithm for loading sparse matrices that were stored in files in the ABHSF. The algorithm works in cases where both the same or different configurations for storing and loading procedures were used. The presented method for different configurations is general. It can be used when a different number of processes, different matrix-processes mapping, and/or a different in-memory storage format is used. Due to this approach, the algorithm reads all stored files by all processes, which results in lower loading times (when compared with the same configuration case). When the configurations are different but determined, it might be possible to develop algorithms adapted especially for such configuration pairs. Such a development represents subjects for future research.

\subsection*{Acknowledgements}

This work was supported by the Czech Science Foundation under Grant No. P202/12/2011.
This work was supported by the IT4Innovations Centre of Excellence project (CZ.1.05/1.1.00/02.0070), funded by the European Regional Development Fund and the national budget of the Czech Republic via the Research and Development for Innovations Operational Programme, as well as Czech Ministry of Education, Youth and Sports via the project Large Research, Development and Innovations Infrastructures (LM2011033).

\bibliographystyle{abbrv}
\bibliography{langr}

\begin{thebibliography}{1}

\bibitem{Barrett:1994}
R.~Barrett, M.~Berry, T.~F. Chan, J.~Demmel, J.~Donato, J.~Dongarra,
  V.~Eijkhout, R.~Pozo, C.~Romine, and H.~V. der Vorst.
\newblock {\em Templates for the Solution of Linear Systems: Building Blocks
  for Iterative Methods}.
\newblock SIAM, Philadelphia, PA, 2nd edition, 1994.

\bibitem{Catalyurek:2010}
U.~V. \c{C}ataly\"{u}rek, C.~Aykanat, and B.~U\c{c}ar.
\newblock On two-dimensional sparse matrix partitioning: Models, methods, and a
  recipe.
\newblock {\em SIAM Journal on Scientific Computing}, 32(2):656--683, 2010.

\bibitem{Langr:2013a}
D.~Langr, I.~{\v S}ime{\v c}ek, and P.~Tvrd{\'\i}k.
\newblock Storing sparse matrices in the adaptive-blocking hierarchical storage
  format.
\newblock In {\em Proceedings of the Federated Conference on Computer Science
  and Information Systems (FedCSIS 2013)}, pages 479--486. IEEE Xplore Digital
  Library, September 2013.

\bibitem{Langr:2013b}
D.~Langr, I.~{\v S}ime{\v c}ek, P.~Tvrd{\'\i}k, and T.~Dytrych.
\newblock Scalable parallel generation of very large sparse matrices.
\newblock In R.~Wyrzykowski, J.~Dongarra, K.~Karczewski, and J.~Waœniewski,
  editors, {\em 10th International Confernce on Parallel Processing and Applied
  Mathematics (PPAM 2013)}, Lecture Notes in Computer Science, pages 178--187.
  Springer Berlin Heidelberg, 2014.
\newblock Accepted for publication.

\bibitem{Langr:2012b}
D.~Langr, I.~{\v S}ime{\v c}ek, P.~Tvrd{\'\i}k, T.~Dytrych, and J.~P. Draayer.
\newblock Adaptive-blocking hierarchical storage format for sparse matrices.
\newblock In {\em Proceedings of the Federated Conference on Computer Science
  and Information Systems (FedCSIS 2012)}, pages 545--551. IEEE Xplore Digital
  Library, September 2012.

\bibitem{Saad:2003}
Y.~Saad.
\newblock {\em Iterative Methods for Sparse Linear Systems}.
\newblock Society for Industrial and Applied Mathematics, Philadelphia, PA,
  USA, 2nd edition, 2003.

\bibitem{HDFGroup:2013}
The {HDF} {G}roup. {H}ierarchical data format version 5, 2000-2013.
\newblock \url{http://www.hdfgroup.org/HDF5/} (accessed June 3, 2013).

\end{thebibliography}

\end{document}